\documentclass[useAMS,usedcolumn,usegraphicx,usenatbib]{mn2e}

\usepackage{amsmath}

\defcitealias{Bell2006}{Paper I}

\title[Conversion factors for specific galaxy types]
      {Molecular line intensities as measures of cloud masses - \\ II. Conversion factors for specific galaxy types}
\author[T.~A.~Bell, S.~Viti and D.~A.~Williams]
{T.~A.~Bell,$^{1,2}$\thanks{E-mail: tab@submm.caltech.edu}
S.~Viti$^{1}$
and D.~A.~Williams$^{1}$\\
$^{1}$Department of Physics \& Astronomy, University College London, Gower Street, London WC1E 6BT\\
$^{2}$Department of Astronomy, California Institute of Technology, Pasadena, CA 91125, USA}

\begin{document}

\date{Accepted 2007 April 4. Received 2007 March 28; in original form 2006 November 29}

\pagerange{\pageref{firstpage}--\pageref{lastpage}} \pubyear{2007}

\maketitle

\label{firstpage}

\begin{abstract}
We present theoretically-established values of the CO-to-H$_{2}$ and C-to-H$_{2}$ conversion factors that may be used to estimate the gas masses of external galaxies. We consider four distinct galaxy types, represented by M51, NGC\,6946, M82 and SMC~N27. The physical parameters that best represent the conditions within the molecular clouds in each of the galaxy types are estimated using a $\chi^{2}$ analysis of several observed atomic fine structure and CO rotational lines. This analysis is explored over a wide range of density, radiation field, extinction, and other relevant parameters. Using these estimated physical conditions in methods that we have previously established, CO-to-H$_{2}$ conversion factors are then computed for CO transitions up to $J=9\to8$. For the conventional CO(1--0) transition, the computed conversion factor varies significantly below and above the canonical value for the Milky Way in the four galaxy types considered. Since atomic carbon emission is now frequently used as a probe of external galaxies, we also present, for the first time, the C-to-H$_{2}$ conversion factor for this emission in the four galaxy types considered.
\end{abstract}



\section{Introduction}\label{Introduction}

The CO-to-H$_{2}$ conversion factor, $X_{\rmn{CO}} = N(\rmn{H_{2}}) / W(\rmn{CO})$ cm$^{-2}$\,(K\,km\,s$^{-1}$)$^{-1}$, where $N(\rmn{H_{2}})$ is the column density of molecular hydrogen (in cm$^{-2}$) and $W(\rmn{CO})$ is the velocity-integrated intensity of the CO $J=1\to0$ rotational line (in K\,km\,s$^{-1}$), is widely used to estimate cloud masses within the Galaxy and in external galaxies. However, as demonstrated in a previous study \citep[][hereafter Paper I]{Bell2006}, the value of $X_{\rmn{CO}}$ is highly sensitive to the local physical conditions, and may vary considerably even within the Galaxy. The appropriate value to use in external galaxies has not so far received theoretical attention. Observationally, studies of external galaxies have inferred values of $X_{\rmn{CO}}$ using a variety of techniques, including virial mass estimates when individual clouds can be resolved \citep{Brouillet1998}, dust continuum observations coupled with some assumptions regarding the gas-to-dust mass ratio \citep*{Boselli2002} and the use of neutral carbon emission as a tracer when its abundance can be determined \citep*{Israel2006}. Such studies confirm the variable nature of $X_{\rmn{CO}}$, which is seen to differ from galaxy to galaxy \citep[e.g.][]{Maloney1988}. The ability to resolve molecular emission from individual clouds within nearby galaxies through interferometric techniques has created new opportunities for studying the interstellar medium in different environments \citep[e.g.][]{Sakamoto1999,Helfer2003}; however, many properties still depend critically on the masses derived for these clouds. The CO-to-H$_{2}$ conversion factor therefore has an increasingly important role to play in advancing our understanding of the properties of interstellar gas in external galaxies. Atomic carbon emission is now also routinely observed in external galaxies and has been proposed as an alternative tracer of molecular mass \citep*[see, e.g.,][]{Papadopoulos2004}, with a C-to-H$_{2}$ conversion factor of the form $X_{\text{C\,{\sc i}}} = N(\rmn{H_{2}}) / W(\text{C\,{\sc i}})$ cm$^{-2}$\,(K\,km\,s$^{-1}$)$^{-1}$, where $W(\text{C\,{\sc i}})$ is the velocity-integrated intensity of the [\mbox{C\,{\sc i}}] 609\,\micron\ fine structure line (in K\,km\,s$^{-1}$).

The purpose of this paper is to determine theoretically from fundamental arguments (as described in \citetalias{Bell2006}) the appropriate values of $X_{\rmn{CO}}$ and $X_{\text{C\,{\sc i}}}$ to use for each of several distinct galaxy types in which the physical conditions are significantly different from each other. We select a sample of nearby galaxies that are representative of the various commonly observed morphological types and determine the physical properties that best describe the neutral interstellar gas within them by fitting observed line intensity ratios for a range of far-infrared (FIR) and submillimetre diagnostic lines. Using the methods of \citetalias{Bell2006}, we then calculate $X_{\rmn{CO}}$ for various rotational transitions of the CO molecule and $X_{\text{C\,{\sc i}}}$ for the 609\,\micron\ fine structure transition of atomic carbon for each of the galaxies selected, taking the physical parameters from the best fitting model in each case.

In Section~\ref{GalaxySample} we discuss the sample of galaxies we have selected to study and introduce them individually. Section~\ref{Fitting} describes our fitting procedure to determine appropriate physical parameters for the interstellar medium (ISM) in each of our sample galaxies. We present the best fit model for each galaxy in Section~\ref{BestFitModels} and discuss the derived physical parameters and the quality of the fits. In Section~\ref{ConversionFactors} we use these models to compute suitable CO-to-H$_{2}$ and C-to-H$_{2}$ conversion factors for each galaxy type considered and we summarize our conclusions in Section~\ref{Conclusions}.


\section{The galaxy sample}\label{GalaxySample}

\begin{table*}
\begin{minipage}{130.2mm}
 \caption{Properties of the four galaxies adopted for this study.}
 \label{GalaxyProperties}
 \begin{tabular}{@{}llllcccccc}
  \hline
  Galaxy & Morphology$^{a}$ & \multicolumn{1}{c}{R.A.(J2000)} & \multicolumn{1}{c}{Dec.(J2000)} & $V_{\rmn{LSR}}$ & Distance & Angular Scale \\
         &                  & \multicolumn{1}{c}{(hh\ mm\ ss)} & \multicolumn{1}{c}{($\degr\ \quad\arcmin\ \quad\arcsec$)} & (km\,s$^{-1}$) & (Mpc) & (pc/arcsec) \\
  \hline
  M51       & SA(s)bc pec & $\,13\ 29\ 52.7$ & $+47\ 11\ 42.0$ & $\phantom{}$464 &    9.7$^{b}$ & $\phantom{}$46.5 \\
  NGC\,6946 & SAB(rs)cd   & $\,20\ 34\ 51.2$ & $+60\ 09\ 17.5$ & $\phantom{4}$55 &    5.5$^{c}$ & $\phantom{}$26.7 \\
  M82       & I0; Sbrst   & $\,09\ 55\ 52.3$ & $+69\ 40\ 45.9$ & $\phantom{}$175 &    3.2$^{d}$ & $\phantom{}$15.5 \\
  SMC N27   & dSB(s)m pec & $\,00\ 48\ 21.1$ & $-73\ 05\ 29.0$ & $\phantom{}$115 & 63~kpc$^{e}$ & $\phantom{1}$0.3 \\
  \hline
 \end{tabular}

 \medskip
 References: $^{a}$\,\citet{deVaucouleurs1991}; $^{b}$\,\citet{Sandage1975}; $^{c}$\,\citet{Tully1988};
 $^{d}$\,\citet{Dumke2001}; $^{e}$\,\citet{Rubio1993}.
\end{minipage}
\end{table*}

To produce CO-to-H$_{2}$ and C-to-H$_{2}$ conversion factors that are applicable to as wide a variety of galaxies as possible, we select four nearby examples that are considered to be prototypical of the main galaxy types, possessing interstellar clouds that span a range of physical conditions and so are likely to require conversion factors that are somewhat different to the canonical value for the Milky Way, $X_{\rmn{CO}}=2\times10^{20}$~cm$^{-2}$\,(K\,km\,s$^{-1}$)$^{-1}$ \citep*{Strong1996,Dame2001}. The examples we consider are the grand-design spiral galaxy M51, the late-type spiral galaxy NGC\,6946, which has a nuclear starburst, the irregular starburst galaxy M82 and nebula N27 within the Small Magellanic Cloud, a dwarf irregular galaxy. Table~\ref{GalaxyProperties} lists the properties of our chosen galaxies; their morphologies \citep[from the RC3 catalogue of][]{deVaucouleurs1991}, the coordinates of their central positions (J2000 epoch), their radial velocities relative to the local standard of rest (LSR), estimated distances (in Mpc) and the corresponding angular scale at that distance (in pc/arcsec). 

We restrict our study to the central positions of these galaxies (with the exception of the Small Magellanic Cloud, see below). This is partly due to the limited amount of relevant observational data in the required transitions. Most studies of the neutral ISM within external galaxies focus on single-beam observations or mapping of the brightest regions, typically their centres, which are known to exhibit strong molecular concentrations. These studies are further restricted due to the large telescope beam sizes involved (typically $>$10 arcsec, although interferometric mapping can reach angular resolutions of $\sim$1 arcsec); these beams can encompass whole ensembles of clouds that exhibit a range of properties. Determination of the physical conditions within these regions is then necessarily limited to deriving average properties for an ensemble of clouds. In this section, the four galaxies in our sample are introduced and observational studies of the interstellar gas within them are summarized.

\subsection{M51}

The grand-design spiral of M51 (NGC\,5194) appears nearly face-on ($i\sim20\degr$) at a distance of 9.7~Mpc (see Table~\ref{GalaxyProperties}), and is seen to be interacting with its smaller companion, NGC\,5195, which lies 4.5~arcmin to the north. It has a metallicity slightly above Solar, determined near its centre to be $12+\log{(\rmn{O/H})}=9.28\pm0.11$ \citep*{Zaritsky1994}. M51 has been classed a Seyfert 2 galaxy by \citet*{Ho1997}, with a central AGN surrounded by a $\sim$100~pc disc of warm and dense gas \citep{Matsushita1998}. This galaxy has been studied extensively and a large amount of observational data are available, including an extended map of [\mbox{C\,{\sc ii}}] 158\,\micron\ emission, taken with the \textit{Kuiper Airborne Observatory} \citep{Nikola2001} and additional observations of FIR fine-structure lines toward the galaxy centre \citep{Negishi2001} using the Long Wavelength Spectrometer (LWS) onboard the \textit{Infrared Space Observatory} (\textit{ISO}). Ground-based single-dish studies mapping the low-lying transitions of $^{12}$CO and $^{13}$CO show the CO to be tightly confined to the spiral arms \citep*[e.g.][]{Garcia-Burillo1993,Nakai1994}. Maps of higher CO transitions, $J=3\to2$ and $4\to3$, have been obtained with the Heinrich Hertz Telescope (HHT) by \citet{Nieten1999} and \citet{Dumke2001}, showing warm molecular gas to be present out to a galactocentric radius of 5~kpc or more within these arms. Recently, the [\mbox{C\,{\sc i}}] 609\,\micron\ fine structure line has also been observed towards the galaxy centre and several positions along the spiral arms \citep{Gerin2000,Israel2002,Kramer2005}, which show that the neutral carbon emission is also extended. Aperture synthesis maps were obtained by \citet{Sakamoto1999} and \citet{Regan2001} at resolutions of 4--6~arcsec in the CO(1--0) line and, more recently, \citet{Matsushita2004} have mapped the inner region in CO(3--2).

\subsection{NGC\,6946}

This nearby low-inclination ($i\sim30\degr$) galaxy, classified as a spiral with a weak bar (SAB) in the RC3 catalogue \citep{deVaucouleurs1991}, is at a distance of 5.5~Mpc (see Table~\ref{GalaxyProperties}) and has an abundance gradient $d\log{(\rmn{O/H})}/dR = -0.089\pm0.003$~dex\,kpc$^{-1}$ \citep{Belley1992}, similar to that of the Milky Way, and a metallicity of $12+\log{(\rmn{O/H})}=9.13\pm0.15$ near its centre \citep{Zaritsky1994}. It has a nuclear starburst \citep[see, e.g.,][]{Engelbracht1996} and a small inner bar, seen clearly in recent interferometric maps of CO(1--0) and CO(2--1) \citep{Schinnerer2006}, that is driving an inflow of molecular gas into the central region of intense star formation. The galaxy shows a pronounced open spiral pattern, the north-eastern arm being the brightest at all wavelengths, indicating a site of vigorous ongoing star formation. One of the first galaxies to be mapped in CO emission \citep{Morris1978}, NGC\,6946 has been relatively well studied in the lower CO transitions, both towards its central region \citep*[e.g.][]{Weliachew1988} and towards its spiral arms \citep{Casoli1990}. More recently, maps of the central region have been made in the $J=2\to1$, $3\to2$ and $4\to3$ transitions of CO and in the [\mbox{C\,{\sc i}}] 609\,\micron\ line \citep{Mauersberger1999,Israel2001}. All these observations show that the gas is predominantly molecular within 5 kpc of the centre and that CO emission is rather extended over the galaxy, but is most concentrated in the central region.

\subsection{M82}

The irregular starburst galaxy M82 (NGC\,3034) is perhaps the most studied of all nearby galaxies due to its spectacular display of active star formation, which dominates its evolution and is thought to have been triggered about 100~Myr ago by tidal interaction with its companion galaxy M81. Originating in the nucleus, the starburst activity within M82 is currently seen to be propagating into the molecular rings, disrupting the surrounding ISM and causing the insterstellar clouds to fragment. The gravitational interaction is also likely to have produced the stellar bar seen in the central $\sim$400~pc \citep{Telesco1991}. As a result of its vigorous star formation, M82 is also one of the most luminous nearby galaxies in the infrared, with $L_{\rmn{IR}}\sim4\times10^{10}~L_{\odot}$. The galaxy has been the subject of many observational studies, including extensive mapping in both low and high (up to $J=7\to6$) transitions of CO \citep*[e.g.][]{Sutton1983,White1994,Mao2000,Seaquist2006}, comprehensive FIR spectroscopy \citep{Colbert1999} and mapping in a variety of molecular species, including HCO$^{+}$ and HCN \citep{Seaquist2000}, SiO \citep{Garcia-Burillo2001} and CO$^{+}$ \citep{Fuente2006}. These observations demonstrate the presence of PDR chemistry over large regions of the galaxy centre and strong, extended CO emission arising from a warm and diffuse ISM.

\subsection{Nebula N27 in the Small Magellanic Cloud}

At just under 63~kpc away, the Small Magellanic Cloud (SMC; NGC\,292) is one of the nearest examples of dwarf irregular galaxy, presenting an excellent opportunity to study the properties of interstellar gas under conditions very different to those in the Milky Way. Its metal content is known to be particularly low, with an oxygen abundance of $12+\log{[\rmn{O/H}]}=7.96\pm0.11$ \citep{Vermeij2002} implying a metallicity of about one tenth Solar, and it possesses a gas-rich interstellar medium that is subjected to strong FUV fluxes due to the presence of a large population of young massive stars \citep{Westerlund1990}. As part of a three-galaxy system that includes the Large Magellanic Cloud (LMC) and the Milky Way Galaxy, the SMC has experienced a disruptive history; an encounter with the LMC some 200~Myr ago is likely to be responsible for the current burst of star-forming activity \citep{Gardiner1996}. Such rapid star formation in a low metallicity environment is also thought to be representative of the conditions within primordial galaxies. In the absence of an obvious central feature, we have chosen to adopt the well-studied region surrounding the N27 nebula as our example of the typical environment within the SMC. The N27 nebula is a bright and relatively compact \mbox{H\,{\sc ii}} region located in the south-western bar of the SMC and is associated with the IRAS point source LIRS 49 \citep{Schwering1989}. It has been mapped in the $J=1\to0$ and $J=2\to1$ lines of CO by \citet*{Rubio1993} and observed in the $J=3\to2$ CO line and transitions of CS, HCO$^{+}$ and HCN by \citet*{Heikkila1999}. More recently, the [\mbox{C\,{\sc i}}] 609\,\micron\ line has been detected in this region for the first time \citep{Bolatto2000} and the FIR fine structure lines have been observed using the \textit{ISO} LWS.


\section{Physical parameters for the galaxies}\label{Fitting}

The atomic fine structure and CO rotational emission lines serve as excellent diagnostics of the physical conditions within PDRs, since they are the dominant cooling lines within the warm gas at the surfaces of molecular clouds. Their range of critical densities and energy level spacings mean that, when used in combination, the intensities of these lines can be used to constrain the density, incident FUV flux and surface temperature of the gas \citep[see, e.g.,][]{Kaufman1999}.

In order to derive appropriate physical parameters for the galaxies in our sample, we compare emission line intensity ratios predicted by PDR models to those detected in each of the four example galaxies. The best fitting models are determined by performing a $\chi^{2}$ goodness of fit analysis of the predicted and observed line ratios for a grid of PDR models covering a large region of parameter space. Section~\ref{ChiSquaredFits} describes the grid of models used for comparison with the galaxies in our sample, the calculation of appropriate line ratios from the observed line intensities and the calculation of $\chi^{2}$ values, from which the best fit models are determined. The observational data obtained from the literature and the resulting line ratios adopted for each galaxy are discussed and the corresponding best fit models are presented in Section~\ref{BestFitModels}. These models are then used in Section~\ref{ConversionFactors} to derive suitable CO-to-H$_{2}$ and C-to-H$_{2}$ conversion factors, $X_{\rmn{CO}}$ and $X_{\text{C\,{\sc i}}}$, for each galaxy type considered.


\subsection{Best fit PDR models using $\bmath{\chi^{2}}$ analysis}\label{ChiSquaredFits}

The short computation times required by the \textsc{ucl\_pdr} code make it ideally suited to the calculation of large grids of models covering substantial regions of parameter space. The models predict the integrated intensities of the fine structure and CO rotational lines emitted by a PDR with the adopted physical parameters. Each model produces line intensities as a function of total cloud depth (at visual extinctions of $0<A_{V}<10$~mag) for cloud ages of $10^{4}$, $10^{5}$, $10^{6}$, $10^{7}$ and $10^{8}$~yr. The standard version of the \textsc{ucl\_pdr} code was used to calculate the grid of models to which the $\chi^{2}$ analysis is then applied. The code is described in detail in \citetalias{Bell2006}. The gas is initially assumed to be fully atomic in the models, with all carbon in C$^{+}$, and the evolving chemistry and thermal balance are then computed for subsequent time-steps. The PDR is treated as a one-dimensional semi-infinite slab of uniform density. We adopt the standard chemical network used in the \textsc{ucl\_pdr} code, neglecting freeze-out of atoms and molecules on to grains. The metallicity-dependence of the gas-phase elemental abundances, dust grain and PAH number densities, photoelectric heating rate and H$_{2}$ formation rate on grains are as described in \citetalias{Bell2006}. Table~\ref{ModelGrid} lists the range of parameter values covered by the grid of models. Homogeneous clouds with hydrogen nuclei number densities of $10^{2} \le n \le 10^{5}$~cm$^{-3}$, subject to incident FUV fluxes of $10 \le G_{0} \le 10^{5}$ times the standard \citet{Habing1968} ISRF have been computed at metallicities $Z/Z_{\odot}$ of 1, 0.5 and 0.1, and cosmic-ray ionization rates $\zeta$ of 5$\times$$10^{-17}$, 5$\times$$10^{-16}$ and 5$\times$$10^{-15}$~s$^{-1}$. A turbulent velocity of $v_{\rmn{turb}}=1.5$~km\,s$^{-1}$ is adopted in all the models, typical of individual Galactic molecular clouds \citep[e.g.][]{Hollenbach1999}. We do not consider the variation of this parameter in our attempts to fit the observed emission from each galaxy, since the line ratios are believed to be fairly insensitive to small changes in the turbulent velocity \citep*{Wolfire1989}. Furthermore, few estimates exist for the degree of turbulent motion within molecular clouds of other galaxies, since observations of extragalactic emission lines do not generally resolve the velocities of individual clouds.

\begin{table}
\begin{center}
 \begin{minipage}{74.7mm}
  \caption{The region of parameter space covered by the grid of PDR models.}
  \label{ModelGrid}
  \begin{tabular}{@{}l r@{$\,\le\,$}c@{$\,\le\,$}l}
   \hline
   Parameter & \multicolumn{3}{l}{Range of values} \\
   \hline
   Cloud density (cm$^{-3}$)  & $10^{2}$ &   $n$   & $10^{5}$ \\
   Incident FUV flux (Habing) & $10^{1}$ & $G_{0}$ & $10^{5}$ \\
   Age of the cloud (yr)      & $10^{4}$ &   $t$   & $10^{8}$ \\
   Cloud size (mag)           & $0$      & $A_{V}$ & $10$ \\
   Metallicity                & \multicolumn{3}{l}{$Z/Z_{\odot}=1.0,0.5,0.1$} \\
   C.R.~ionization rate (s$^{-1}$) & \multicolumn{3}{l}{$\zeta=5\times,50\times,500$$\times$$10^{-17}$} \\
   Turbulent velocity (km\,s$^{-1}$) & \multicolumn{3}{l}{$v_{\rmn{turb}}=1.5$} \\
   \hline
  \end{tabular}
 \end{minipage}
\end{center}
\end{table}

Comparing PDR model predictions to extragalactic observations is never straightforward. Individual molecular clouds are rarely resolved in single-dish observations, particularly so in the case of FIR fine structure lines, which have typically been observed with the \textit{ISO} LWS with a beam size of $\sim$70--80~arcsec. Several phases of the ISM are therefore contained within a single beam and the line intensity ratios will represent the combined emission from these phases. The physical properties derived from comparison with PDR models are then the \textit{average} properties for the molecular clouds within the region, although they may be biased towards those of the brighter components. By using line ratios, rather than fitting the line intensities directly, the beam area filling factors of the two emission lines are cancelled out, assuming that they come from the same clouds.

In our fitting of the observed fine structure and CO rotational lines, we will consider the following line intensity ratios:
\begin{enumerate}
 \renewcommand{\theenumi}{(\arabic{enumi})}
  \item {[\mbox{O\,{\sc i}}]}  63\,\micron$/$[\mbox{C\,{\sc ii}}]$_{\rmn{PDR}}$
  \item {[\mbox{O\,{\sc i}}]} 145\,\micron$/$[\mbox{O\,{\sc i}}] 63\,\micron
  \item {[\mbox{C\,{\sc ii}}]}$_{\rmn{PDR}}/$[\mbox{C\,{\sc i}}] 609\,\micron
  \item {[\mbox{C\,{\sc ii}}]}$_{\rmn{PDR}}/$CO(3--2)
  \item {[\mbox{C\,{\sc i}}]} 609\,\micron$/$CO(3--2)
  \item {CO(3--2)}$/$CO(1--0)
\end{enumerate}
when the necessary line intensities are available in the literature. The calculation of [\mbox{C\,{\sc ii}}]$_{\rmn{PDR}}$ is described below.

In order to derive meaningful ratios, the various line intensities must share a common spatial resolution. Given that the lines are observed with different beam sizes, it is necessary to convolve the integrated intensities to the spatial resolution appropriate to the largest beam size, typically that of the \textit{ISO} LWS, which has an average half-power beam width (HPBW) of $\sim$75~arcsec. When observational data at the necessary resolution are not available in the literature, we perform the convolution using the scaling factor:
\begin{equation}\label{Eqn:Convolution}
 \Phi = \sqrt{\frac{\theta_{\rmn{b},i}^{2} + \theta_{\rmn{s},x}^{2}}
                   {\theta_{\rmn{b},f}^{2} + \theta_{\rmn{s},x}^{2}}}
 \times \sqrt{\frac{\theta_{\rmn{b},i}^{2} + \theta_{\rmn{s},y}^{2}}
                   {\theta_{\rmn{b},f}^{2} + \theta_{\rmn{s},y}^{2}}}\,,
\end{equation}
where $\theta_{\rmn{b},i}$ and $\theta_{\rmn{b},f}$ are the initial (observed) and final beam sizes (HPBW in arcsec), and $\theta_{\rmn{s},x}$ and $\theta_{\rmn{s},y}$ are the source sizes along the major and minor axes (FWHM in arcsec). Of course, if the source is circular then these two axes will have the same dimensions. The observed intensity is then multiplied by this factor to obtain the intensity at the desired resolution.

Since [\mbox{C\,{\sc ii}}] 158\,\micron\ emission arises from both ionized and neutral gas, the fraction originating from the ionized medium must be subtracted before a comparison with PDR models can be made. We have chosen to estimate the fraction of [\mbox{C\,{\sc ii}}] emission from the ionized medium using the observed [\mbox{N\,{\sc ii}}] 122\,\micron\ fine structure line, in the manner of \citet{Malhotra2001} and \citet{Contursi2002}. The [\mbox{N\,{\sc ii}}] emission occurs exclusively in the ionized gas, so by assuming a constant ratio between the [\mbox{C\,{\sc ii}}] and [\mbox{N\,{\sc ii}}] line intensities in the diffuse ionized medium (DIM), taken to be the dominant ionized component, the contribution of ionized gas to [\mbox{C\,{\sc ii}}] emission can be estimated. The [\mbox{C\,{\sc ii}}]/[\mbox{N\,{\sc ii}}] ratio derived by these authors is 4.3, based upon observations of ionized gas in the Milky Way and using Solar abundances for carbon and nitrogen. In reality, this ratio is likely to vary from region to region and the metallicities of the galaxies studied here are known to differ. However, the abundance ratio has been shown to be independent of metallicity in both normal and irregular galaxies \citep{Garnett1999} and in view of the lack of more reliable estimates of the ionized gas contribution, the use of this ratio is justified. The intensity of [\mbox{C\,{\sc ii}}] emission originating from PDRs is then
\begin{equation}\label{Eqn:IonizedComponent}
 I_{\text{[C\,{\sc ii}]}_{\rmn{PDR}}} = I_{\text{[C\,{\sc ii}]}} - 4.3\times I_{\text{[N\,{\sc ii}]}} \qquad [\rmn{erg\,s^{-1}\,cm^{-2}\,sr^{-1}}],
\end{equation}
where $I_{\text{[C\,{\sc ii}]}}$ and $I_{\text{[N\,{\sc ii}]}}$ are the observed intensities of the [\mbox{C\,{\sc ii}}] 158\,\micron\ and [\mbox{N\,{\sc ii}}] 122\,\micron\ lines.

Finally, the [\mbox{C\,{\sc i}}] and CO line intensities are usually measured as velocity-integrated antenna temperatures (in K\,km\,s$^{-1}$). These values must therefore be converted to integrated intensities (erg\,s$^{-1}$\,cm$^{-2}$\,sr$^{-1}$) before they can be compared to the FIR fine structure lines and model predictions. This is achieved using the relation:
\begin{equation}
 I = \frac{2k\nu^{3}}{c^{3}} \int T_{\rmn{A}}\,\rmn{d}v \qquad [\rmn{erg\,s^{-1}\,cm^{-2}\,sr^{-1}}],
\end{equation}
where $\nu$ is the transition frequency (in Hz) and $\int T_{\rmn{A}}\,\rmn{d}v$ is the antenna temperature integrated over velocity (in K\,km\,s$^{-1}$).

Once the relevant line intensity ratios have been determined for a particular galaxy, we calculate the reduced $\chi^{2}$ goodness of fit for each model in the grid,
\begin{equation}\label{Eqn:ChiSquared}
 \chi^{2} = \frac{1}{(N_{\rmn{obs}} - N_{\rmn{par}})} \sum_{i} \left( \frac{R_{i}^{\rmn{obs}} - R_{i}^{\rmn{mod}}}{\sigma_{i}} \right)^{2},
\end{equation}
where $N_{\rmn{obs}}$ is the number of \textit{independent} line ratios compared to observations, $N_{\rmn{par}}$ is the number of free parameters in the models and, for each line ratio $i$ considered, $R_{i}^{\rmn{obs}}$ is the observed value, $R_{i}^{\rmn{mod}}$ is the model value and $\sigma_{i}$ is the estimated error on the observed value. We take the typical observational error to be $\sim$20~per cent of the line intensity, including calibration uncertainties, leading to a combined error of $\sim$30~per cent on the line ratios. In addition to these uncertainties, errors introduced by convolving the intensities to the required resolution and by correcting for the ionized gas contribution also play a significant role in determining the value of $\chi^{2}$. These errors are hard to quantify and have therefore been neglected here, but their inclusion would serve to generally decrease the value of $\chi^{2}$ obtained for each model. Uncertainties in the contribution of the ionized gas emission would only affect line ratios involving the [\mbox{C\,{\sc ii}}] 158\,\micron\ line, reducing the significance of these ratios in constraining the best fit model. Note that three of the line ratios are linked and so not independent, [\mbox{C\,{\sc ii}}]$_{\rmn{PDR}}/$[\mbox{C\,{\sc i}}] 609\,\micron, [\mbox{C\,{\sc i}}] 609\,\micron$/$CO(3--2) and [\mbox{C\,{\sc ii}}]$_{\rmn{PDR}}/$CO(3--2), thus reducing the number of independent line ratios $N_{\rmn{obs}}$ by one.

The best fit model for the physical conditions in each galaxy is therefore the model with the minimum reduced $\chi^{2}$ value. In the next section we list the observed and predicted line ratios for each galaxy in our sample and present contour maps of reduced $\chi^{2}$ across the grid of models for each galaxy. The physical parameters derived from the best fit models are discussed for each of the galaxies in our sample.


\section{Adopted line ratios and best fit model parameters}\label{BestFitModels}

\begin{table*}
\begin{minipage}{112.5mm}
 \caption{Observed line ratios for each galaxy in the sample.}
 \label{LineRatios1}
 \begin{tabular}{@{}lcccccc}
  \hline
  Galaxy & $\frac{\mbox{[O\,{\sc i}]\ 63}}{\mbox{[C\,{\sc ii}]}_{\rmn{PDR}}}$
         & $\frac{\mbox{[O\,{\sc i}]\ 145}}{\mbox{[O\,{\sc i}]\ 63}}$
         & $\frac{\mbox{[C\,{\sc ii}]}_{\rmn{PDR}}}{\mbox{[C\,{\sc i}]\ 609}}$
         & $\frac{\mbox{[C\,{\sc ii}]}_{\rmn{PDR}}}{\mbox{CO(3--2)}}$
         & $\frac{\mbox{[C\,{\sc i}]\ 609}}{\mbox{CO(3--2)}}$
         & $\frac{\mbox{CO(3--2)}}{\mbox{CO(1--0)}}$ \\
  \hline
  M51       & 0.730 &  n/a  & $\phantom{1}$39.4 & $\phantom{6}$43.9 & 1.114 & 14.74 \\
  NGC\,6946 & 0.935 & 0.058 & $\phantom{}$112.2 & $\phantom{6}$58.2 & 0.519 & 14.11 \\
  M82       & 1.329 & 0.092 & $\phantom{}$128.2 & $\phantom{}$605.6 & 4.726 & 16.11 \\
  SMC N27   & 0.873 & 0.026 & $\phantom{}$219.7 & 512.3$^{\star}$ & 2.331$^{\star}$ & 8.694$^{\star}$ \\
  \hline
 \end{tabular}

 \medskip
 $^{\star}$CO(3--2) data were not available at the location of the N27 nebula so these line
 ratios were instead calculated using the CO(2--1) integrated intensity derived from the
 observational data of \citet{Rubio1993}.
\end{minipage}
\end{table*}

\begin{table*}
\begin{minipage}{112.5mm}
 \caption{Line ratios predicted by the best fit model for each galaxy in the sample.}
 \label{LineRatios2}
 \begin{tabular}{@{}lcccccc}
  \hline
  Galaxy & $\frac{\mbox{[O\,{\sc i}]\ 63}}{\mbox{[C\,{\sc ii}]}_{\rmn{PDR}}}$
         & $\frac{\mbox{[O\,{\sc i}]\ 145}}{\mbox{[O\,{\sc i}]\ 63}}$
         & $\frac{\mbox{[C\,{\sc ii}]}_{\rmn{PDR}}}{\mbox{[C\,{\sc i}]\ 609}}$
         & $\frac{\mbox{[C\,{\sc ii}]}_{\rmn{PDR}}}{\mbox{CO(3--2)}}$
         & $\frac{\mbox{[C\,{\sc i}]\ 609}}{\mbox{CO(3--2)}}$
         & $\frac{\mbox{CO(3--2)}}{\mbox{CO(1--0)}}$ \\
  \hline
  M51       & 0.728 & 0.037 & $\phantom{1}$49.7 & $\phantom{6}$54.0 & 1.087 & 21.84 \\
  NGC\,6946 & 0.885 & 0.036 & $\phantom{1}$49.2 & $\phantom{6}$45.7 & 0.928 & 23.42 \\
  M82       & 2.081 & 0.042 & $\phantom{1}$85.4 & $\phantom{}$500.1 & 5.854 & 14.03 \\
  SMC N27   & 0.488 & 0.021 & $\phantom{}$209.4 & 1113$^{\star}$ & 5.315$^{\star}$ & 6.218$^{\star}$ \\
  \hline
 \end{tabular}

 \medskip
 $^{\star}$Calculated using the CO(2--1) integrated line intensity to allow comparison with the
 observed line ratios.
\end{minipage}
\end{table*}

The line intensities used to derive the relevant line ratios are taken from the literature, applying the convolution formula in equation~\ref{Eqn:Convolution} when necessary in order to obtain a set of integrated intensities at a common resolution for each source. All the FIR data included have been acquired using the \textit{ISO} LWS, with a typical HPBW of 70--80~arcsec \citep{Lloyd2003}, corresponding to spatial scales of $\sim$1--3.5~kpc at the distances of these nearby galaxies. All other observations at higher spatial resolution have therefore been convolved to a beam size of 75~arcsec ($\Omega_{\rmn{b}}=1.1\times10^{-7}$~sr).

Table~\ref{LineRatios1} lists the adopted line intensity ratios for each galaxy, to which the predicted line ratios from each model are compared. The majority of the FIR fine structure line intensities are taken from \citet{Negishi2001}, who present \textit{ISO} LWS observations for a large sample of galaxies. The line fluxes listed in table~2 of their paper have been converted to intensities using the extended source correction factors of \citet{Lloyd2003} and an effective aperture of $\Omega_{\rmn{b}}=1.1\times10^{-7}$~sr. For M51, the CO(1--0), CO(3--2) and [\mbox{C\,{\sc i}}] 609\,\micron\ integrated intensities listed in table~4 of \citet{Kramer2005} have been adopted, since these values are already at an appropriate angular resolution of 80~arcsec, derived by smoothing available map data. The CO and [\mbox{C\,{\sc i}}] data for NGC\,6946 are taken from \citet{Israel2001,Israel2002} and have been convolved to the required resolution using an assumed source size of $26\,\rmn{arcsec}\times23\,\rmn{arcsec}$ \citep{Nieten1999}. The adopted line intensities for M82 were drawn from observational data presented by \citet{Colbert1999}, \citet{Mao2000} and \citet{Israel2002} for the FIR, CO and [\mbox{C\,{\sc i}}] transitions, respectively. \citet{Bolatto2000} have used line intensities observed toward nebula N27 of the Small Magellanic Cloud to estimate the physical conditions within the region by comparison with the PDR models of \citet{Kaufman1999}. The \textit{ISO} LWS data presented in their paper are used here, adjusted to our adopted beam size of 75~arcsec, along with their observed value for the [\mbox{C\,{\sc i}}] integrated intensity in the region. The CO(1--0) map of \citet{Rubio1993} covers a smaller area within N27, but is thought to encompass most of the CO emission \citep{Bolatto2000} and is therefore adopted here. Unfortunately, no CO(3--2) data are currently available in the literature for this source. \citet{Rubio1993} have mapped the CO(2--1) line in the region and we therefore use these data to calculate the ratios [\mbox{C\,{\sc ii}}]$_{\rmn{PDR}}/$CO(2--1), [\mbox{C\,{\sc i}}] 609\,\micron$/$CO(2--1) and CO(2--1)$/$CO(1--0) in place of the equivalent CO(3--2) line ratios. The CO and [\mbox{C\,{\sc i}}] data were convolved to the appropriate angular resolution assuming a source size of $110\,\rmn{arcsec}\times110\,\rmn{arcsec}$, determined from the maps of \citet{Rubio1993}. The [\mbox{N\,{\sc ii}}] 122\,\micron\ data used to estimate the contribution of the ionized medium to the [\mbox{C\,{\sc ii}}] 158\,\micron\ emission are taken from \citet{Kramer2005}, \citet{Negishi2001} and \citet{Colbert1999} for the relevant regions in M51, NGC\,6946 and M82, respectively. The resulting contributions from the ionized gas in these galaxies are found to range from $\sim$30--40 per cent of the total [\mbox{C\,{\sc ii}}] emission, the greatest correction being applied to the M51 data. No [\mbox{N\,{\sc ii}}] data are available in the literature for the region of interest within the SMC.

\begin{table*}
\begin{minipage}{122.3mm}
 \caption{Best fit physical parameters for each of the four galaxies, based on the model with the lowest $\chi^{2}$ value.}
 \label{BestFitParameters}
 \begin{tabular}{@{}lccccccc}
  \hline
  Galaxy & $n$ (cm$^{-3}$) & $G_{0}$ (Habing) & $t$ (yr) & $A_{V}$ (mag) & $Z/Z_{\odot}$ & $\zeta$ (s$^{-1}$) & $\chi^{2}_{\rmn{min}}$ \\
  \hline
  M51       & $7^{_{+11\phantom{.}}}_{^{-3}}\times10^{3}$ & $20^{_{+13}}_{^{^{-14}}}$
            & $10^{7}$ & 9.8 &   2 & 5$\times$$10^{-17}$ & 1.6 \\

  NGC\,6946 & $1^{_{+1.8}}_{^{-0.3}}\times10^{4}$ & $20^{_{+38}}_{^{^{-15}}}$
            & $10^{7}$ & 9.9 &   3 & 5$\times$$10^{-17}$ & 6.0 \\

  M82       & $1^{_{+0.9}}_{^{-0.1}}\times10^{3}$ & $500^{_{+96\phantom{6}}}_{^{^{-92}}}$
            & $10^{8}$ & 5.3 &   1 & 5$\times$$10^{-17}$ & 2.9 \\

  SMC N27   & $1^{_{+4.2}}_{^{-0.4}}\times10^{3}$ & $100^{_{+356}}_{^{^{-61}}}$
            & $10^{5}$ & 2.8 & 0.1 & 5$\times$$10^{-17}$ & 4.9 \\
  \hline
 \end{tabular}
\end{minipage}
\end{table*}

\begin{figure}
 \includegraphics[width=84mm]{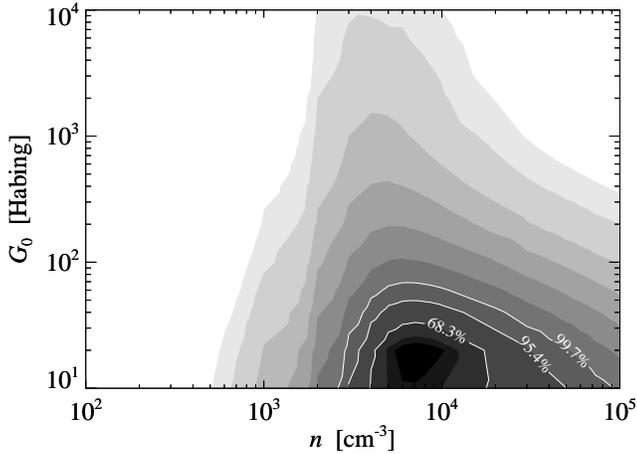}
 \caption{Centre of M51. Contours of the reduced $\chi^{2}$ fit to the data are shown across the range of parameter space covered by the grid of models in the density--radiation ($n$--$G_{0}$) plane. The cloud age and size are fixed at the best fit values for M51 listed in Table~\ref{BestFitParameters}. Contour levels are 2.2 (black), 2.7, 4.0 ($1\sigma$ confidence level), 7.8 ($2\sigma$), 13.5 ($3\sigma$), 25, 50, 100, 200, 500, 1000 and $>$1000 (white).}
 \label{ReducedChi:M51}
\end{figure}

\begin{figure}
 \includegraphics[width=84mm]{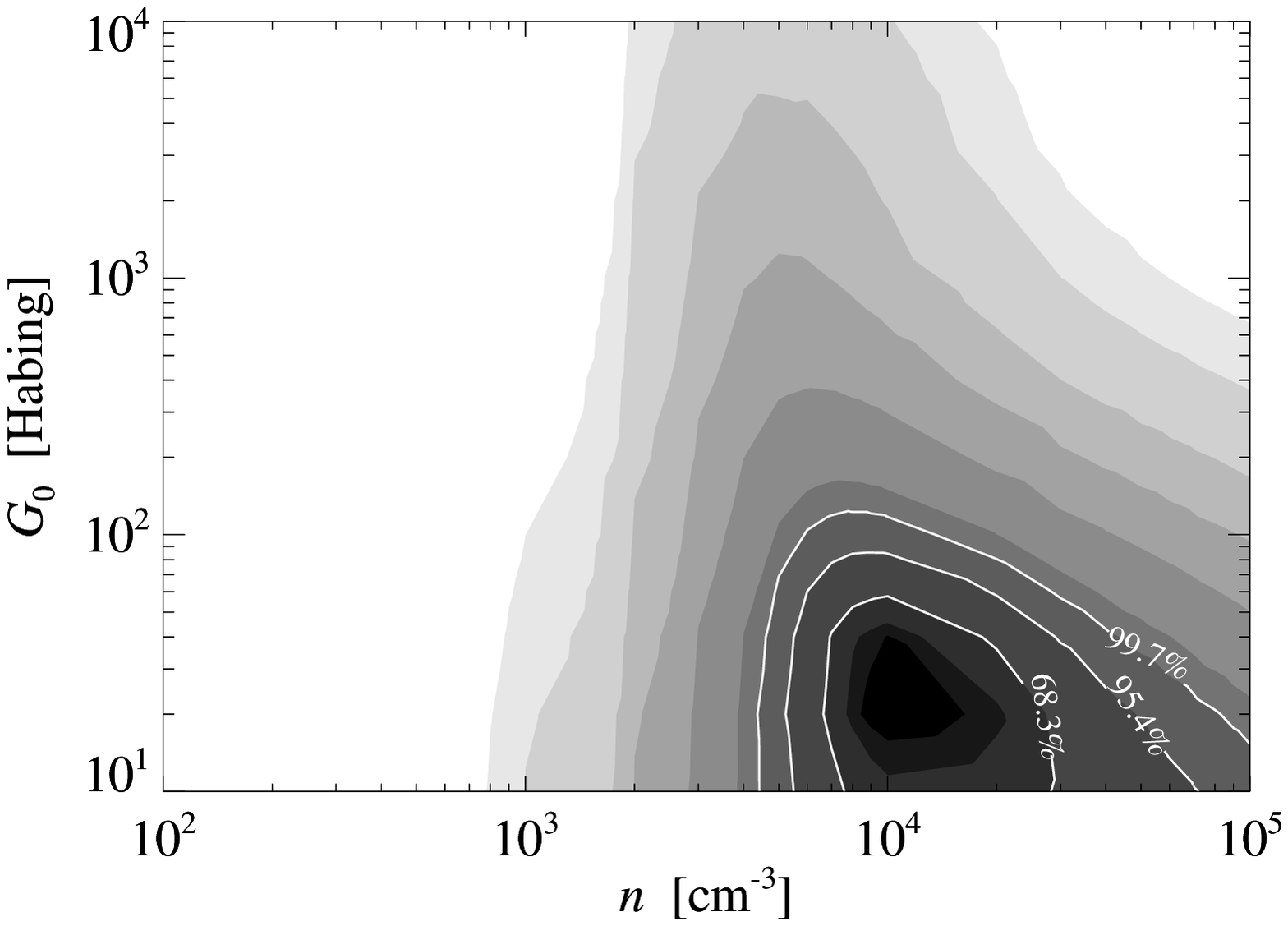}
 \caption{Centre of NGC\,6946. Contours of the reduced $\chi^{2}$ fit to the data are shown across the range of parameter space covered by the grid of models in the density--radiation ($n$--$G_{0}$) plane. The cloud age and size are fixed at the best fit values for NGC\,6946 listed in Table~\ref{BestFitParameters}. Contour levels are 7.9 (black), 8.4, 9.7 ($1\sigma$ confidence level), 13.6 ($2\sigma$), 19.2 ($3\sigma$), 25, 50, 100, 200, 500, 1000 and $>1000$ (white).}
 \label{ReducedChi:NGC6946}
\end{figure}

The line ratios predicted by the best fit model for each galaxy are listed in Table~\ref{LineRatios2} and the physical parameters of each best fit model are listed in Table~\ref{BestFitParameters}. Fig.~\ref{ReducedChi:M51}--\ref{ReducedChi:SMC} show contours of reduced $\chi^{2}$, calculated for each grid model across the adopted range of densities and FUV fluxes, for each galaxy in our sample. White labelled contours mark the 1, 2 and $3\sigma$ confidence levels for each $\chi^{2}$ distribution. The cloud age and size are fixed in each plot to the best fit values listed in Table~\ref{BestFitParameters}. Models with Solar metallicity were initially used to fit the observed line ratios of M51, NGC\,6946 and M82, which display similar metallicities. When determining the best fit models for the SMC region, the analysis was restricted to only those models with $Z/Z_{\odot}=0.1$, appropriate for this low metallicity environment.

\subsection{M51}

For the central region of M51, in which our adopted beam size encompasses an area approximately 3.5~kpc across, we find that the observed line intensity ratios are best fit by average conditions characterised by reasonably dense ($n=7\times10^{3}$~cm$^{-3}$) gas immersed in an ambient FUV field ($G_{0}=20$ Habing). These values agree remarkably well with those determined by \citet{Kramer2005}, who perform a similar $\chi^{2}$ fit to the observed line ratios using the PDR models presented by \citet{Kaufman1999}. They find that local densities of $n\sim10^{4}$~cm$^{-3}$ and FUV fluxes of $G_{0}\sim20$ Habing best fit the observations, both at the central position and within the spiral arms of the galaxy. This level of agreement is encouraging, suggesting that the method is reliable and that our models are consistent with those of others. However, our approach has the additional benefit of determining the average cloud size and estimated degree of chemical evolution within the region, since the models produce line intensities that are both depth- and time-dependent. We find that the line ratios are best fit by cloud models with visual extinctions of $A_{V}\sim10$~mag and ages of $\sim$10~Myr (when evolving from initially atomic gas). Such conditions are expected in dense regions where many clouds along the line of sight give rise to optically thick emission (with the possible exception of the [\mbox{C\,{\sc i}}] and [\mbox{C\,{\sc ii}}] lines) and the gas is maintained in molecular form. Our best fit model predicts a cosmic-ray ionization rate of $\zeta=5$$\times$$10^{-17}$~s$^{-1}$, consistent with the low radiation field and quiescent star formation activity in this region. The fact that \citet{Kramer2005} find similar conditions in the spiral arms of M51 suggests that our models might be equally applicable to these regions.

The centre of M51 is known to have a metallicity slightly above Solar \citep[see, e.g.,][]{Zaritsky1994}. To see if a better fit to the observed line ratios might be obtained by using a higher metallicity, we computed additional models with the same density, incident FUV flux and cosmic-ray ionization rate as the best fit model, but with metallicities of $Z=2\,Z_{\odot}$, $3\,Z_{\odot}$, $4\,Z_{\odot}$ and $5\,Z_{\odot}$ (the original best fit model was computed at Solar metallicity). The resulting reduced $\chi^{2}$ values for these models are 1.62, 1.63, 1.73 and 1.74 for the $Z=2\,Z_{\odot}$ to $5\,Z_{\odot}$ cases, respectively. Models with $Z=2\,Z_{\odot}$ and $Z=3\,Z_{\odot}$ show a slight improvement in the goodness of fit, with reduced $\chi^{2}$ values lower than that of the original best fit model ($\chi^{2}_{\rmn{min}}=1.68$), but given the uncertainties in the line strengths, these differences are unlikely to be significant. It is probable that these line ratios are only able to provide loose constraints on the metallicity. Since the $Z=2\,Z_{\odot}$ model produces the best fit, we adopt this metallicity for the central region of M51, however, we note that, in the absence of additional constraints, the uncertainty in the metallicity will lead to an uncertainty in the derived value of $X_{\rmn{CO}}$. The minimum reduced $\chi^{2}$ for this model is a factor of 2 better than that achieved by \citet{Kramer2005}. We therefore regard this model as representative of the conditions within interstellar molecular gas in normal spiral galaxies.

\subsection{NGC\,6946}

Similar physical parameters are found to best fit the line ratios determined for the central region of NGC\,6946 ($n=10^{4}$~cm$^{-3}$, $G_{0}=20$ Habing, $t\sim10$~Myr, $A_{V}\sim10$~mag and $\zeta=5$$\times$$10^{-17}$~s$^{-1}$). This is surprising, given that the galaxy possesses a starburst nucleus and higher FUV fluxes would therefore be expected. We note that the fit is not as good here as it was for M51, with a minimum reduced $\chi^{2}$ of 7.41 for the best fit model and some degree of uncertainty in the appropriate value of $G_{0}$, as demonstrated by the elongated contours in Fig.~\ref{ReducedChi:NGC6946}. Since the metallicity in the central region of NGC\,6946 is above Solar \citep{Zaritsky1994}, we consider models with higher metallicities in the same manner as for M51. We find that increasing the metallicity to $Z=3\,Z_{\odot}$ produces a better fit ($\chi^{2}_{\rmn{min}}=5.99$), whilst higher metallicities yield poorer fits ($\chi^{2}>6$). We therefore adopt $Z=3\,Z_{\odot}$ as the best fit model for the central region of NGC\,6946. Again, we note that the metallicity is only weakly constrained by this technique and the choice of appropriate value is somewhat uncertain.

\citet{Contursi2002} fitted the line intensity ratio [\mbox{C\,{\sc ii}}]$_{\rmn{PDR}}/$[\mbox{O\,{\sc i}}] 63\,\micron, together with the line-to-FIR continuum ratio, ([\mbox{C\,{\sc ii}}]$_{\rmn{PDR}} + [$\mbox{O\,{\sc i}}] 63\,\micron)$/$FIR, for \textit{ISO} LWS observations of various positions within NGC\,6946, again employing the PDR model predictions of \citet{Kaufman1999}. At the central position, they derived for the neutral gas an average density of $n\approx2\times10^{3}$~cm$^{-3}$ and an incident FUV flux of $G_{0}\approx900$ Habing. These values are significantly different to those that we have derived, predicting the gas to be a factor of 5 lower in density and, more dramatically, the incident FUV flux to be close to a factor of 50 higher. Such a large difference suggests that, by comparing different line ratios, we may be probing different gas components within the region. The large beam size, corresponding to an area $\sim$2~kpc across at the distance of NGC\,6946, makes it likely that the emission arises from a mixture of active and less active regions. The FIR continuum traces the strength of the FUV flux, which is absorbed by the dust and re-emitted in the IR. However, whilst this emission is optically thin, and therefore seen across the whole region, the FUV flux responsible is rapidly attenuated in the clouds it is incident upon, so the surface area of the gas that is actually subjected to the strong FUV flux may constitute only a small fraction of the total area. By using this diagnostic to derive the physical conditions, the inferred cloud properties may therefore be biased towards those of only a small fraction of the total gas content. Both sets of physical parameters are likely to represent the conditions within certain regions of the galaxy and we therefore consider both when deriving conversion factors in the next section.

\begin{figure}
 \includegraphics[width=84mm]{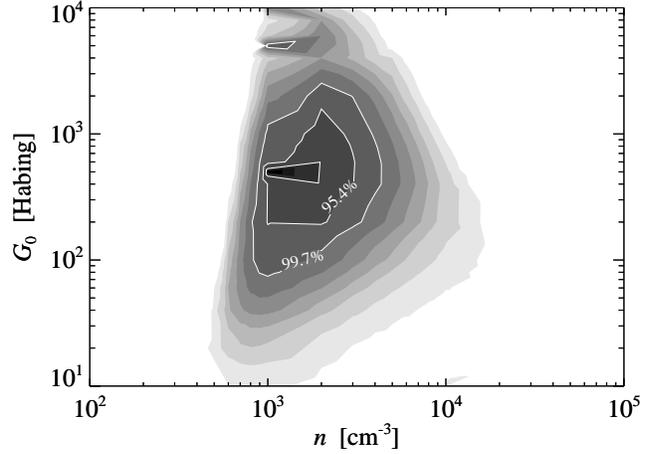}
 \caption{Centre of M82. Contours of the reduced $\chi^{2}$ fit to the data are shown across the range of parameter space covered by the grid of models in the density--radiation ($n$--$G_{0}$) plane. The cloud age and size are fixed at the best fit values for M82 listed in Table~\ref{BestFitParameters}. Contour levels are 3.4 (black), 3.9, 5.2 ($1\sigma$ confidence level), 9.1 ($2\sigma$), 14.7 ($3\sigma$), 20, 25, 30, 35, 42, 50 and $>50$ (white).}
 \label{ReducedChi:M82}
\end{figure}

\subsection{M82}

\begin{figure}
 \includegraphics[width=84mm]{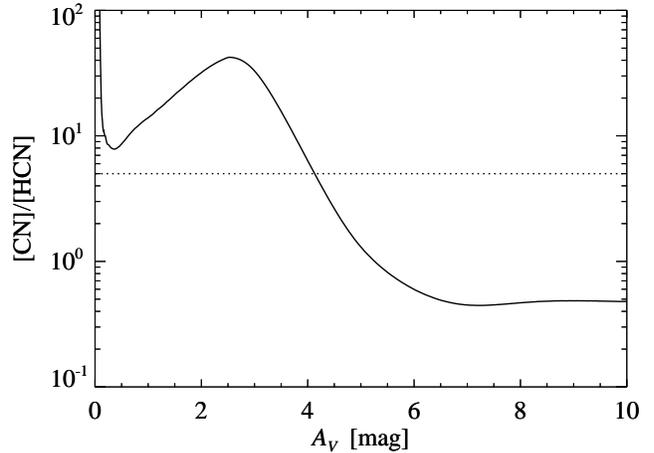}
 \caption{Centre of M82. The abundance ratio of CN to HCN ([CN]$/$[HCN]; solid line) as a function of visual extinction $A_{V}$ for the best fit model ($n=10^{3}$~cm$^{-3}$, $G_{0}=500$~Habing), compared to the ratio observed by \citet[][dotted line]{Fuente2005}.}
 \label{M82abundances}
\end{figure}

The starburst galaxy M82 is found to have average gas properties of $n=10^{3}$~cm$^{-3}$ and $G_{0}=500$ Habing within its central region (an area $\sim$1~kpc across is enclosed in the 75~arcsec beam at this distance), according to the best fit model determined by our $\chi^{2}$ analysis ($\chi^{2}_{\rmn{min}}=2.9$). These properties are consistent with those derived by \citet{Colbert1999}, who find $n\approx2000$~cm$^{-3}$ and $G_{0}\approx600$ Habing using a combination of \mbox{H\,{\sc ii}} and PDR models to self-consistently fit the observed FIR ionic and atomic fine structure lines. A typical cloud age of $\sim$100~Myr is predicted by the best fit model. We find that the observed line ratios are best fit by smaller clouds, with total visual extinctions of $A_{V}\approx5$~mag. This is in agreement with the recent observational study of \citet{Fuente2005}, who find that the observed abundance ratio $[\rmn{CN}]/[\rmn{HCN}]\ga5$, determined for the central 650~pc of M82, requires cloud sizes of $A_{V}<5$--6~mag. For comparison, we plot the $[\rmn{CN}]/[\rmn{HCN}]$ abundance ratio produced by our best fit model as a function of visual extinction in Fig.~\ref{M82abundances}. From this plot, it can be seen that a cloud size of $A_{V}\la4$~mag is also required to maintain an abundance ratio $>5$ in this model.

\citet*{Suchkov1993} proposed an increased cosmic-ray ionization rate of $\zeta=4\times10^{-15}$~s$^{-1}$ in the centre of M82 to account for the observed physical conditions of the molecular gas. The grid models with the same density and incident FUV flux as the best fit model, but with cosmic-ray ionization rates of $\zeta=5\times10^{-16}$ and $5\times10^{-15}$~s$^{-1}$ (i.e.~10 and 100 times the best fit rate) produce reduced $\chi^{2}$ values of 8.7 and 5.5, respectively. Both are higher than the best fit model value of $\chi^{2}_{\rmn{min}}=2.9$, although the model with a rate 100 times higher shows a better fit than the model with only a factor 10 increase in the cosmic-ray ionization rate. Since no improvement in the goodness of fit is obtained by using a higher value of $\zeta$, we adopt the physical parameters of the original best fit model when determining appropriate conversion factors for starburst regions.

\begin{figure}
 \includegraphics[width=84mm]{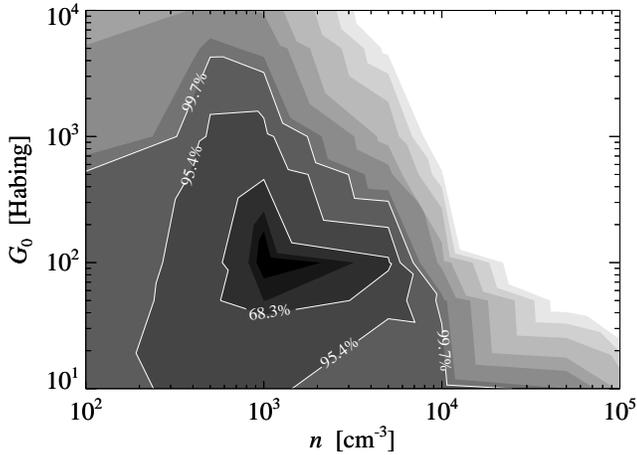}
 \caption{N27 region of the Small Magellanic Cloud. Contours of the reduced $\chi^{2}$ fit to the data are shown across the range of parameter space covered by the grid of models in the density--radiation ($n$--$G_{0}$) plane. The cloud age and size are fixed at the best fit values for the region listed in Table~\ref{BestFitParameters}. Contour levels are 5.4 (black), 5.9, 7.2 ($1\sigma$ confidence level), 11.1 ($2\sigma$), 16.7 ($3\sigma$), 20, 30, 40, 60, 80, 100 and $>100$ (white).}
 \label{ReducedChi:SMC}
\end{figure}

\subsection{SMC (N27)}

In the low metallicity environment of the Small Magellanic Cloud, the molecular gas is known to exhibit significantly different properties to those typically found in regions with metallicities closer to Solar. By restricting our parameter search to models with metallicity $Z=0.1\,Z_{\odot}$ (the average value derived for the SMC by \citealt{Dufour1984}) we obtain a best fit model for the line ratios observed in the SMC nebula N27 that predicts the presence of clouds with reasonably dense gas ($n=10^{3}$~cm$^{-3}$) subjected to elevated FUV fluxes ($G_{0}=100$~Habing), likely due to the active star formation and reduced extinction in this low metallicity environment. These results are in very good agreement with those obtained by \citet{Bolatto2000}, determined using FIR fine structure lines and continuum emission to be $n\sim300$--1000~cm$^{-3}$ and $G_{0}\sim30$--100~Habing. Furthermore, whilst previous models have had to impose the limit of low $A_{V}$ observed in these clouds, this property is independently predicted as a result of our fitting procedure ($A_{V}=2.8$~mag in our best fit model). The line ratios observed in N27 are less well fit than those of M51 and M82 ($\chi^{2}_{\rmn{min}}=4.9$ for the best fit model of N27). An analysis of the agreement between model predictions and observations shows that the predicted [\mbox{C\,{\sc ii}}]$/$CO(2--1) ratio is the most discrepant of the six line ratios considered, being over a factor of 2 higher than the observed value. We note that in our adopted line ratios for N27 we have not accounted for the ionized gas contribution to the [\mbox{C\,{\sc ii}}] 158\,\micron\ emission, since the required [\mbox{N\,{\sc ii}}] 122\,\micron\ data are unavailable in the literature. However, subtracting the ionized gas contribution would decrease the observed value of the [\mbox{C\,{\sc ii}}]$/$CO(2--1) ratio, which would actually \textit{reduce} the quality of the fit. Instead, a more favourable explanation for the disagreement between predicted and observed line ratios is the high O/C abundance ratio in the SMC, a factor 2.5 greater than the local value \citep{Wilson2005}. This was not taken into account in the models. We adopt the physical parameters of the best fit model for our consideration of suitable conversion factors for dwarf irregular galaxies.

\subsection{Discussion of best fit parameters and their uncertainties}

The best fit ages determined for each galaxy provide an indication of the degree of chemical evolution from atomic gas. Whilst ages of $10^{7}$--$10^{8}$~yr are predicted for M51, NGC\,6946 and M82, suggesting that the emission arises from predominantly molecular gas, a relatively young age of $10^{5}$~yr is derived for the SMC, which implies that the gas in this region is mainly atomic, either due to the slower formation rate of H$_{2}$ in the low metallicity environment or because the reduced shielding results in greater photodissociation of molecular species. The radiation field is stronger still in M82, yet the molecular gas is predicted to be well-evolved. This is probably because M82 has a higher metallicity and a higher typical $A_{V}$ than are found in the SMC, and both of these promote chemical evolution. The degree of chemical evolution can have a strong effect on the CO-to-H$_{2}$ conversion factor, reducing the value of $X_{\rmn{CO}}$ at early times \citepalias[see][for details]{Bell2006}. The cosmic-ray ionization rate for each best fit model is found to take its standard value in all cases, suggesting that the line ratios considered here may not be sensitive to this parameter. Whilst the $\chi^{2}$ analysis indicates that models with metallicities above Solar may be more appropriate for the centres of M51 and NGC\,6946, these metallicities cannot be tightly constrained with our choice of line ratios, as discussed in previous sections.

To provide some estimate of the uncertainty in the best fit density and radiation field strength for each galaxy, 1, 2 and $3\sigma$ confidence level contours are overlaid on the plots of reduced $\chi^{2}$ in Fig.~\ref{ReducedChi:M51}--\ref{ReducedChi:SMC}, and the range of each parameter contained in the $1\sigma$ region is taken as the uncertainty in that parameter, as listed in Table~\ref{BestFitParameters}. The uncertainty in the best fit gas density is typically found to be a factor of 2--3, although for region N27 in the SMC the density is less well constrained and may be up to a factor of 5 higher. The estimated FUX fluxes suffer a slightly greater degree of uncertainty, generally being constrained to within a factor of 4. Again, the greatest degree of uncertainty in $G_{0}$ is found for the SMC, consistent with the relatively poor fit that was obtained for this galaxy. Given the weak constraints on the metallicity provided by our fitting procedure, we also adopt an uncertainty of a factor of 2 in metallicity. The estimated uncertainties in these parameters will be used in Section~\ref{Conclusions} to examine the variation in derived conversion factors.

In the next section we use the best fit models for each galaxy in our sample to derive CO-to-H$_{2}$ and C-to-H$_{2}$ conversion factors that are appropriate for molecular gas within these environments and suggest transitions of CO that might yield the most reliable mass estimate in each case.


\section{Appropriate conversion factors for molecular mass estimates in different galaxy types}\label{ConversionFactors}

Having determined the physical conditions that best represent the average properties of the neutral ISM within each galaxy type, we now derive CO-to-H$_{2}$ and C-to-H$_{2}$ conversion factors appropriate for those conditions in the manner described in \citetalias{Bell2006}. In addition to the standard value of $X_{\rmn{CO}}$, relevant to CO $J=1\to0$ emission, we extend our study to include conversion factors for higher rotational lines of CO and for the [\mbox{C\,{\sc i}}] 609\,\micron\ line that might serve as useful alternatives under certain conditions. The \textsc{ucl\_pdr} code calculates the integrated intensities of all CO rotational lines up to $J=11\to10$ and values of $X_{\rmn{CO}}$ can therefore be derived for each transition using a more general definition of the CO-to-H$_{2}$ ratio,
\begin{equation}\label{Eqn:XFactor2}
 X_{\rmn{CO}}^{u\to l} = \frac{N(\rmn{H_{2}})}{\int T_{\rmn{A}}(\rmn{CO}\ J=u\to l)\,\rmn{d}v} \quad [\rmn{cm^{-2}\,(K\,km\,s^{-1})^{-1}}],
\end{equation}
where $N(\rmn{H_{2}})$ is the column density of H$_{2}$, as before, and $T_{\rmn{A}}(\rmn{CO}\ J=u\to l)$ is the antenna temperature of the CO $J=u\to l$ line. We will use the notation $X_{\rmn{CO}}^{u\to l}$ to distinguish between the different CO-to-H$_{2}$ conversion factors throughout this section. The C-to-H$_{2}$ conversion factor $X_{\text{C\,{\sc i}}}$ for each galaxy can be derived in a similar way, using the [\mbox{C\,{\sc i}}] 609\,\micron\ integrated intensities calculated by the \textsc{ucl\_pdr} code.

\begin{figure}
 \includegraphics[width=84mm]{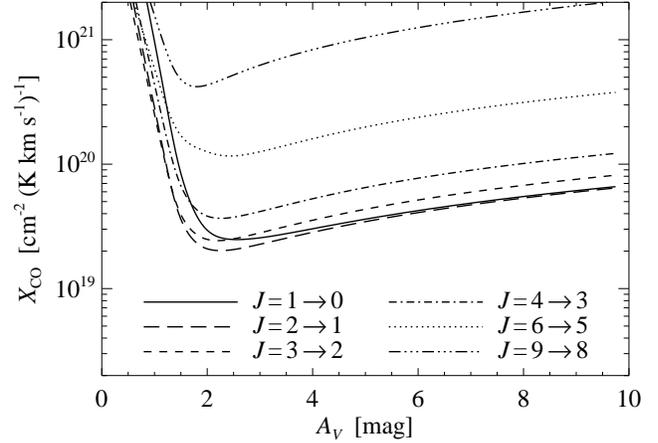}
 \caption{Normal spiral galaxies. $X_{\rmn{CO}}$ versus $A_{V}$ profiles for the best fit physical parameters for the centre of M51. The values of $X_{\rmn{CO}}$ shown are for emission from the various rotational transitions of CO.}
 \label{Xfactor:M51}
\end{figure}

\begin{figure}
 \includegraphics[width=84mm]{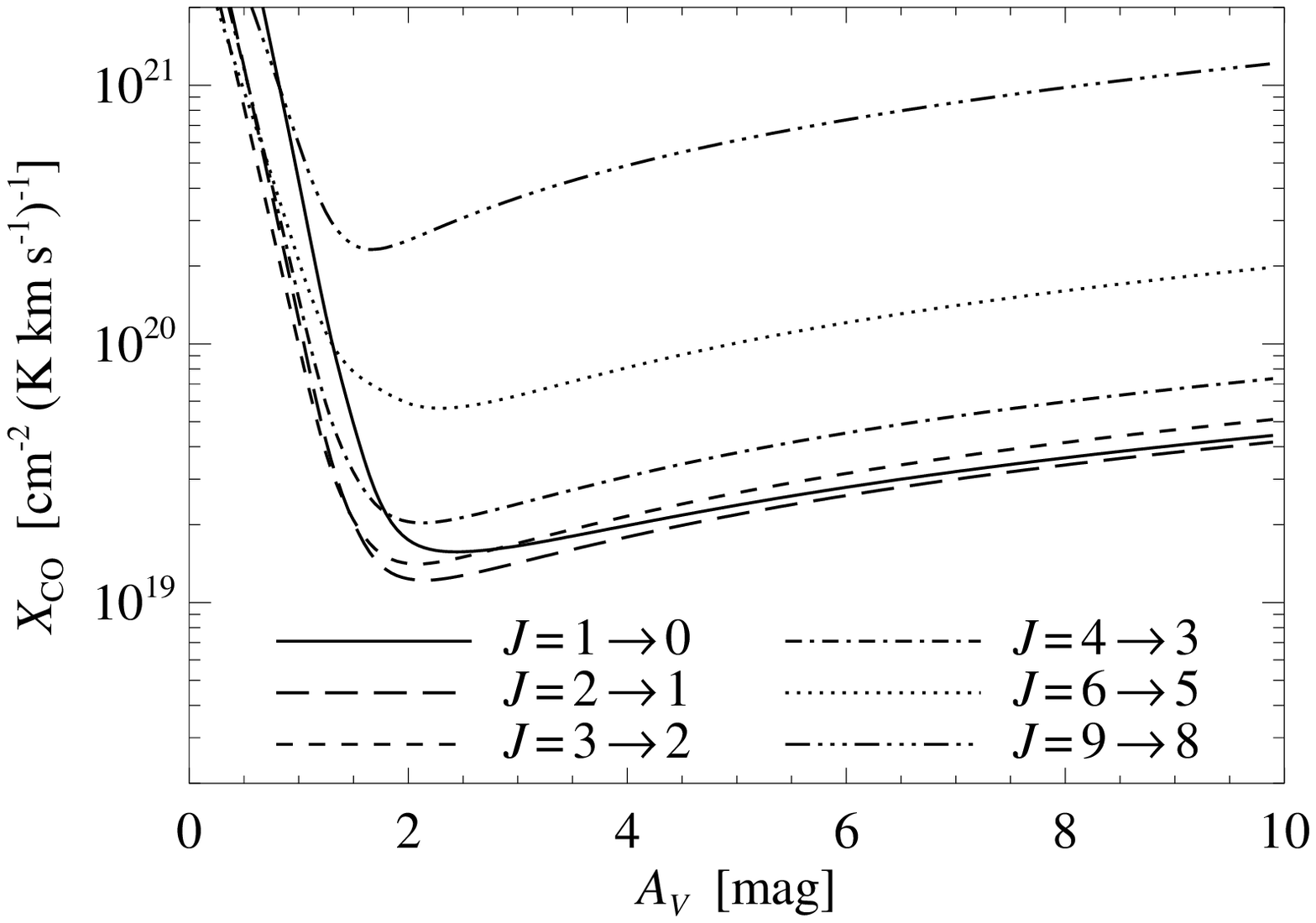}
 \caption{Weak starbust nuclei. $X_{\rmn{CO}}$ versus $A_{V}$ profiles for the best fit physical parameters for the centre of NGC\,6946. Values of $X_{\rmn{CO}}$ for the various rotational transitions of CO are shown.}
 \label{Xfactor:NGC6946a}
\end{figure}

Fig.~\ref{Xfactor:M51}--\ref{Xfactor:SMC} show $X_{\rmn{CO}}$ as a function of visual extinction $A_{V}$ for various transitions of CO, from the best fit model for each galaxy in our sample. Recall that, for these $X_{\rmn{CO}}$ profiles, the observationally-relevant conversion factor in each transition corresponds to the profile minimum, which represents the peak emission from the cloud (or ensemble of clouds; see \citetalias{Bell2006}). Fig.~\ref{CarbonXfactors} shows $X_{\text{C\,{\sc i}}}$ as a function of $A_{V}$ for each galaxy. Again, the profile minima represent the conversion factors relevant to observations. Table~\ref{Xfactors} lists the minimum values of $X_{\rmn{CO}}$ and $X_{\text{C\,{\sc i}}}$ [in units of $10^{20}$ cm$^{-2}$\,(K\,km\,s$^{-1}$)$^{-1}$] for each galaxy type in various transitions of CO (1--0, 2--1, 3--2, 4--3, 6--5 and 9--8), together with the cloud depths at which they occur (in $A_{V}$). These are therefore the appropriate values of $X_{\rmn{CO}}$ and $X_{\text{C\,{\sc i}}}$ that should be used to derive molecular mass estimates in these regions. The results for each of the four galaxy types are discussed briefly below. For the remainder of this section, the units of the CO-to-H$_{2}$ and C-to-H$_{2}$ conversion factors, cm$^{-2}$\,(K\,km\,s$^{-1}$)$^{-1}$, are omitted to save space.

\subsection{Normal spiral galaxies}

We adopted the conditions within the central $\sim$3~kpc of M51 as characteristic of those within normal late-type spiral galaxies, both in their centres and in their spiral arms. Using the best fit model for these conditions, we find a value for the CO(1--0) conversion factor of $X_{\rmn{CO}}^{1\to0}=2.5\times10^{19}$. This value is an order of magnitude \textit{lower} than the canonical CO-to-H$_{2}$ conversion factor derived for the Solar neighbourhood, $X_{\rmn{CO}}^{1\to0}=2\times10^{20}$ \citep{Dame2001}. The drop in $X_{\rmn{CO}}$ values is due to both the increased density and higher metallicity, which result in stronger CO emission \citepalias[see][for details]{Bell2006}. Observational studies of the centres of nearby spiral galaxies have consistently inferred values of $X_{\rmn{CO}}^{1\to0}$ that are lower than the canonical value \citep[see, e.g.,][and references therein]{Maloney1988,Sorai2000,Israel2003}, in agreement with our theoretical value. Furthermore, \citet{Israel2006} have recently determined the CO-to-H$_{2}$ conversion factor for the centre of M51 to be $X_{\rmn{CO}}^{1\to0}=5.0\pm2.5\times10^{19}$ from observations of neutral carbon and CO emission within the region. This is consistent with the value obtained from our best fit model and we feel that this independent observational verification shows our theoretical approach to be reliable. Looking to the higher transition lines of CO, the CO(2--1), CO(3--2) and CO(4--3) lines all give similar conversion factors (2.0$\times10^{19}$, 2.4$\times10^{19}$ and 3.7$\times10^{19}$, respectively) and become optically thick ($\tau>1$) beyond $A_{V}\approx2$~mag. The CO(6--5) and CO(9--8) lines remain optically thin to greater cloud depths, but are fairly weak due to the low gas temperatures ($\sim$150--10~K from the surface to the cloud interior). The lower transitions are therefore more likely to yield reliable molecular mass estimates, particularly when several lines intensities are used in combination. The appropriate C-to-H$_{2}$ conversion factor, determined from the best fit model, is $X_{\text{C\,{\sc i}}}=5.9\times10^{19}$ and its variation with visual extinction displays a similar trend to that of the CO(1--0) conversion factor, though the minimum occurs nearer the cloud surface (see Fig.~\ref{CarbonXfactors}).

\subsection{Starburst nuclei}

Our best fit physical parameters for the centre of NGC\,6946 are similar to those obtained for M51, representative of more quiescent interstellar gas, although a slightly higher density ($10^{4}$~cm$^{-3}$ instead of $7\times10^{3}$~cm$^{-3}$) and higher metallicity ($3\,Z_{\odot}$ instead of $2\,Z_{\odot}$) are predicted. As such, the derived CO-to-H$_{2}$ conversion factors are also similar, though they are typically 20--50~per cent lower than those found for M51 (see Fig.~\ref{Xfactor:NGC6946a} and Table~\ref{Xfactors}). The best fit model produces a CO(1--0) conversion factor $X_{\rmn{CO}}^{1\to0}=1.6\times10^{19}$ and a C-to-H$_{2}$ conversion factor $X_{\text{C\,{\sc i}}}=3.5\times10^{19}$. \citet{Israel2001} derived a value of $X_{\rmn{CO}}^{1\to0}\sim1\times10^{19}$ for the centre of NGC\,6946, which is remarkably similar to our theoretical value. The depth-dependent profile of $X_{\text{C\,{\sc i}}}$ is similar to that of $X_{\rmn{CO}}$ for the lower CO transitions, as was the case for M51.

\begin{figure}
 \includegraphics[width=84mm]{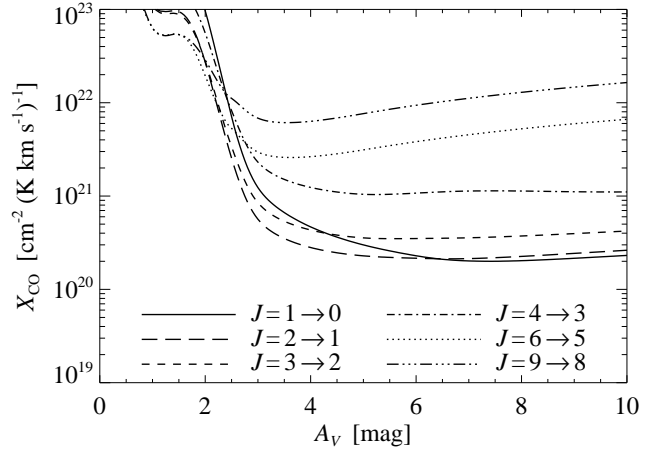}
 \caption{Starburst nuclei. $X_{\rmn{CO}}$ versus $A_{V}$ profiles for the physical parameters derived by \citet{Contursi2002} for the centre of NGC\,6946. Values of $X_{\rmn{CO}}$ for the various rotational transitions of CO are shown.}
 \label{Xfactor:NGC6946b}
\end{figure}

As discussed in the previous section, these conditions may be selectively sampling a particular component within the galaxy centre and may not be fully representative of typical nuclear starburst activity. We therefore also consider an alternative model for the starburst nucleus of NGC\,6946, based on the parameters determined by \citet{Contursi2002}, i.e.~$n=2\times10^{3}$~cm$^{-3}$ and $G_{0}=10^{3}$ Habing. The resulting $X_{\rmn{CO}}$ profiles for the various CO transitions are shown in Fig.~\ref{Xfactor:NGC6946b} and the appropriate conversion factors to be used with each transition to derive molecular masses are listed in Table~\ref{Xfactors}. The increased FUV radiation striking the interstellar gas in the \citet{Contursi2002} model leads to enhanced photodissociation of CO, pushing the C$^+$/C/CO transition deeper into the cloud, where the lower gas temperatures ($\sim$40~K) result in reduced emission strengths. The CO(1--0) conversion factor is $X_{\rmn{CO}}^{1\to0}=2.0\times10^{20}$, the value obtained for local regions within the Milky Way. From examining Fig.~\ref{Xfactor:NGC6946b}, it can be seen that the CO(4--3) line may prove to be the best tracer of total molecular mass in this environment, since the $X_{\rmn{CO}}^{4\to3}$ conversion factor remains fairly constant to high visual extinction. The appropriate value for the conversion factor in this case is $X_{\rmn{CO}}^{4\to3}=1.0\times10^{21}$. The corresponding C-to-H$_{2}$ conversion factor for the \citet{Contursi2002} model is $X_{\text{C\,{\sc i}}}=1.2\times10^{20}$. Its depth-dependent profile (shown in Fig.~\ref{CarbonXfactors}) displays a more pronounced minimum than that of the CO(4--3) line, making it slightly less reliable as a tracer of total molecular mass under these conditions.

\subsection{Irregular starburst galaxies}

\begin{figure}
 \includegraphics[width=84mm]{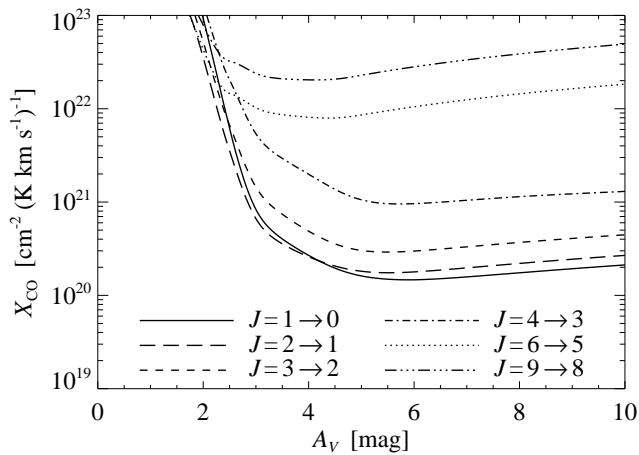}
 \caption{Irregular starburst galaxies. $X_{\rmn{CO}}$ versus $A_{V}$ profiles for the best fit physical parameters for the centre of M82. Values of $X_{\rmn{CO}}$ for the various rotational transitions of CO are shown.}
 \label{Xfactor:M82}
\end{figure}

\citet{Smith1991} have derived the CO-to-H$_{2}$ conversion factor for the central region of M82 from observations of the dust continuum. They find a value of $\approx$$1.2\times10^{20}$, which is in good agreement with the value obtained from our best fit model, $X_{\rmn{CO}}^{1\to0}=1.5\times10^{20}$. The $X_{\rmn{CO}}$ profile minima for the low transitions of CO occur deep within the cloud at $A_{V}\approx6$~mag (see Fig.~\ref{Xfactor:M82}), as in the previous case, due to the higher FUV flux (500~Habing). The conversion factors for transition lines up to CO(4--3) all show fairly constant values for visual extinctions above $\sim$5~mag, suggesting that accurate molecular masses may be obtained for larger cloud sizes. However, the best fit model predicts typical cloud sizes of $A_{V}\la5$~mag for M82, in accordance with observations, so the appropriate CO-to-H$_{2}$ conversion factors for clouds of low $A_{V}$ are likely to vary quite significantly, as shown by the sharp rise in $X_{\rmn{CO}}$ towards the left hand side of Fig.~\ref{Xfactor:M82}. The difficulties of using the CO-to-H$_{2}$ conversion factor in such translucent clouds have been discussed extensively \citep[see, e.g.,][and references therein]{Magnani2003}. The C-to-H$_{2}$ conversion factor, however, is less variable at low visual extinction (see Fig.~\ref{CarbonXfactors}) and its minimum value, $X_{\text{C\,{\sc i}}}=1.3\times10^{20}$, occurs at $A_{V}=4.3$~mag, making it a better tracer of the molecular mass in these clouds.

\subsection{Dwarf irregular galaxies}

\begin{figure}
 \includegraphics[width=84mm]{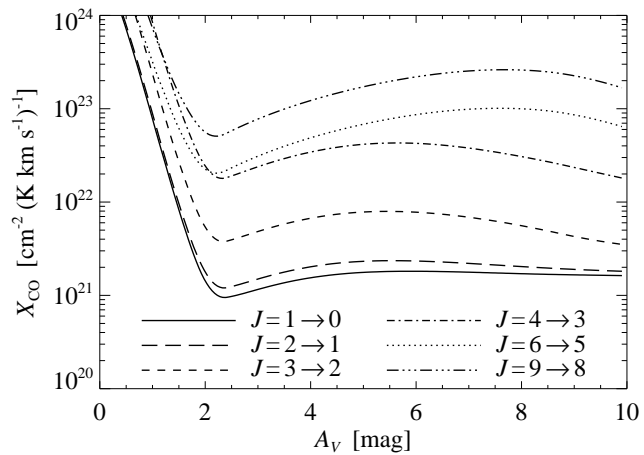}
 \caption{Dwarf irregular galaxies. $X_{\rmn{CO}}$ versus $A_{V}$ profiles for the best fit physical parameters for the N27 nebula of the Small Magellanic Cloud. Values of $X_{\rmn{CO}}$ for the various rotational transitions of CO are shown.}
 \label{Xfactor:SMC}
\end{figure}

The best fit model for the N27 nebula within the Small Magellanic Cloud can provide conversion factors suitable for metal-poor dwarf irregular galaxies. It predicts a CO(1--0) conversion factor of $X_{\rmn{CO}}^{1\to0}=9.5\times10^{20}$, approximately 5 times \textit{greater} than the canonical value for the Milky Way. This is primarily due to the low metallicity ($Z=0.1\,Z_{\odot}$), as discussed in \citetalias{Bell2006}. The low visual extinction in these clouds may mean that the appropriate value is higher still, since $X_{\rmn{CO}}$ increases nearer the cloud surface (see Fig.~\ref{Xfactor:SMC}). \citet{Leroy2006} derive a CO-to-H$_{2}$ conversion factor of $X_{\rmn{CO}}^{1\to0}=6\pm1\times10^{21}$ for regions within the SMC that are occupied by CO, using new FIR continuum maps to determine the dust mass. This is a factor of 6 higher than the theoretical value predicted by our best fit model and may imply that the treatment of metallicity-dependence in the model does not fully account for the reduction in CO emission needed to raise the value of $X_{\rmn{CO}}$ (metallicity is the physical parameter that causes the greatest variation in the value of $X_{\rmn{CO}}$; see \citetalias{Bell2006}). Alternatively, if typical cloud sizes in the SMC are $A_{V}\approx1.6$~mag, rather than the 2.8~mag predicted by the best fit model, then the appropriate CO-to-H$_{2}$ conversion factor would be $X_{\rmn{CO}}^{1\to0}=6\times10^{21}$, due to the steep increase in $X_{\rmn{CO}}$ at low $A_{V}$. Such cloud sizes are indeed found by \citet{Leroy2006}.

\begin{figure}
 \includegraphics[width=84mm]{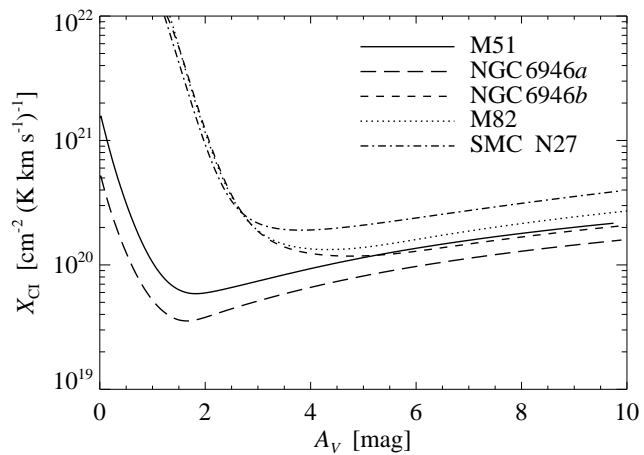}
 \caption{$X_{\text{C\,{\sc i}}}$ versus $A_{V}$ profiles for each galaxy. The profile labelled NGC\,6946\textit{a} is produced by the best fit physical parameters for the centre of NGC\,6946 obtained in this work; the profile labelled NGC\,6946\textit{b} is produced by the physical parameters derived by \citet{Contursi2002}.}
 \label{CarbonXfactors}
\end{figure}

The C-to-H$_{2}$ conversion factor derived from the best fit model is $X_{\text{C\,{\sc i}}}=1.9\times10^{20}$, however, the minimum of the $X_{\text{C\,{\sc i}}}$ profile occurs at $A_{V}\sim4$~mag (see Fig.~\ref{CarbonXfactors}), deeper into the cloud than that of the CO-to-H$_{2}$ profiles, making it less appropriate as a tracer of molecular mass in this case.

\begin{table*}
\begin{minipage}{178mm}
 \caption{Recommended values for the CO-to-H$_{2}$ and C-to-H$_{2}$ conversion factors for each galaxy type. CO-to-H$_{2}$ conversion factors are listed for various transitions of CO. Values of $X_{\rmn{CO}}$ and $X_{\text{C\,{\sc i}}}$ are given in units of $10^{20}$~cm$^{-2}$\,(K\,km\,s$^{-1}$)$^{-1}$.}
 \label{Xfactors}
 \begin{tabular}{@{}l c@{ }rc c@{ }rc c@{ }rc c@{ }rc c@{ }rc c@{ }rc c@{ }c}
  \hline
  Galaxy type & \multicolumn{2}{l}{$J=1\to0$} && \multicolumn{2}{l}{$J=2\to1$} && \multicolumn{2}{l}{$J=3\to2$}
             && \multicolumn{2}{l}{$J=4\to3$} && \multicolumn{2}{l}{$J=6\to5$} && \multicolumn{2}{c}{$J=9\to8$}
             && \multicolumn{2}{l}{[\mbox{C\,{\sc i}}] 609\,\micron} \\
              &  $A_{V}$  &  $X_{\rmn{CO}}$   &&  $A_{V}$  &  $X_{\rmn{CO}}$   &&  $A_{V}$  &  $X_{\rmn{CO}}$
             &&  $A_{V}$  &  $X_{\rmn{CO}}$   &&  $A_{V}$  &  $X_{\rmn{CO}}$   &&  $A_{V}$  &  $X_{\rmn{CO}}$
             &&  $A_{V}$  &  $X_{\text{C\,{\sc i}}}$ \\
  \hline
  Normal spiral
              &   2.6 &   0.3   &&   2.3 &   0.2   &&   2.2 &   0.2
             &&   2.3 &   0.4   &&   2.4 &   1.2   &&   1.8 &   4.2
             &&   1.8 &   0.6   \\
  Starburst nucleus$^{a}$
              &   2.4 &   0.2   &&   2.2 &   0.1   &&   2.1 &   0.1
             &&   2.1 &   0.2   &&   2.3 &   0.6   &&   1.6 &   2.3
             &&   1.7 &   0.4   \\
  Starburst nucleus$^{b}$
              &   7.4 &   2.0   &&   6.7 &   2.1   &&   5.5 &   3.5
             &&   5.2 &  10.4   &&   3.6 &  25.9   &&   3.5 &  61.2
             &&   4.8 &   1.2   \\
  Irregular starburst
              &   5.9 &   1.5   &&   5.5 &   1.7   &&   5.5 &   2.9
             &&   5.7 &   9.5   &&   4.4 &  79.4   &&   4.1 & 203.7
             &&   4.3 &   1.3   \\
  Dwarf irregular
              &   2.4 &   9.5   &&   2.4 &  12.0   &&   2.3 &  37.7
             &&   2.3 & 179.3   &&   2.2 & 203.9   &&   2.2 & 508.0
             &&   3.8 &   1.9   \\
  \hline
 \end{tabular}

 \medskip
 $^{a}$\,Determined for the best fit model obtained for NGC\,6946 in this work (see Section~\ref{BestFitModels}).\newline
 $^{b}$\,Model values calculated for the physical parameters of \citet{Contursi2002}.
\end{minipage}
\end{table*}

The conversion factors listed in Table~\ref{Xfactors} provide a useful reference for observers, presenting `standard' conversion factors that are more appropriate to these diverse environments than the canonical value derived for the Milky Way. By obtaining data in multiple transitions of CO and \mbox{C\,{\sc i}}, several conversion factors can be used to provide better constraints on the estimated molecular mass within a region.


\section{Discussion and Conclusions}\label{Conclusions}

We have determined appropriate physical parameters that represent the average conditions within four key extragalactic environments; the interstellar gas within the spiral arms and nuclei of normal late-type spiral galaxies (modelled on the central region of M51), the starburst nuclei of spiral galaxies and more extreme starburst environments (including the centre of the barred spiral galaxy NGC\,6946 and the prototypical starburst galaxy M82) and the low metallicities and high UV fields typical of dwarf irregular galaxies (in this case, modelled on region N27 of the Small Magellanic Cloud). Best fit models for these environments have been determined by fitting observed line intensity ratios in the selected regions for a range of FIR and sub-mm diagnostic lines.

The resulting models have been used to derive appropriate CO-to-H$_{2}$ and C-to-H$_{2}$ conversion factors that will allow more reliable molecular mass estimates to be inferred from observations of CO rotational lines and the [\mbox{C\,{\sc i}}] 609\,\micron\ fine structure line in a wide variety of extragalactic environments. These conversion factors are listed in Table~\ref{Xfactors}. Conversion factors are presented for a range of CO transitions, enabling estimates of the molecular mass to be further constrained using several line intensities simultaneously. Furthermore, different emission lines may probe specific components within the interstellar medium, e.g.~dense and hot gas in the case of the high $J$ CO transitions. 

Uncertainties in the best fit physical parameters have been determined for each galaxy, and these are now used to examine the corresponding variation in the derived CO-to-H$_{2}$ conversion factors. The uncertainties in the best fit gas density and FUV flux are listed in Table~\ref{BestFitParameters} and we assume a factor of 2 uncertainty in the metallicity. Uncertainties in the other physical parameters are not considered here. $X_{\rmn{CO}}$ decreases with increasing density, as the CO emission becomes stronger nearer the cloud edge, whilst it generally increases with FUV flux, as CO is more readily photodissociated; $X_{\rmn{CO}}$ depends strongly on metallicity, decreasing as metallicity increases and CO emission becomes stronger \citepalias[for a detailed discussion of these trends, see][]{Bell2006}. The sensitivity of the C-to-H$_{2}$ conversion factors to these uncertainties is not discussed here, but is found to be very similar.

For the centre of M51, the uncertainty in the gas density typically causes $X_{\rmn{CO}}$ to vary by less than a factor of 2, with the exception of the higher CO transitions ($J=6\to5$ and above), for which variation by over a factor of 3 is found. Uncertainty in the best fit FUV flux has little or no influence on $X_{\rmn{CO}}$, with at most a factor of 2 variation in the $J=6\to5$ and $9\to8$ conversion factors. The assumed uncertainty of a factor of 2 in metallicity has the greatest impact on the derived values of $X_{\rmn{CO}}$, causing them to vary by up to a factor of 4. The higher transition lines of CO are more sensitive to variations in the gas density and temperature, since their critical densities for excitation are higher and greater temperatures are required to excite them. For the best fit parameters for NGC\,6946, the uncertainty in the density leads to variations in $X_{\rmn{CO}}$ of up to a factor of 2, though without the increase in the higher transition lines that was found for M51. The slightly larger FUV flux uncertainty leads to a marginal increase in the degree of variation of $X_{\rmn{CO}}$, though it is still typically no more than a factor of 2. Variation in the metallicity causes up to a factor of 4 variation in $X_{\rmn{CO}}$.

Relatively tight constraints on the best fit density and radiation field strength for M82 mean that the resulting variation in $X_{\rmn{CO}}$ is also small and always less than a factor of 2. Assuming that the metallicity is again uncertain by a factor of 2, larger variations in $X_{\rmn{CO}}$ are produced than were seen in M51 or NGC\,6946 ($X_{\rmn{CO}}$ varies by a factor of 3--4 in the lower transitions and by over 6 in the $6\to5$ and $9\to8$ transitions). The larger variation with metallicity seen in M82 is probably due to the elevated FUV flux, which, combined with the change in shielding that accompanies the change in metallicity, can enhance the photodissociation of CO.

The best fit physical parameters determined for the SMC suffer the greatest uncertainties and this is reflected in the degree of variation in the conversion factors. The low metallicity adopted for this galaxy ($Z=0.1\,Z_{\odot}$) also causes $X_{\rmn{CO}}$ to be more sensitive to change in the physical parameters. Therefore, the large uncertainty in the density (up to a factor of 5) leads to variations in $X_{\rmn{CO}}$ that range from a factor of 4 for the $J=1\to0$ transition to over a factor of 35 for the $9\to8$ transition line. The weak dependence of the CO-to-H$_{2}$ conversion factor on the FUV flux means that the uncertainty in this parameter produces only a factor of 2 variation in $X_{\rmn{CO}}$. Uncertainty in the metallicity causes $X_{\rmn{CO}}$ to vary by a factor of 4--5.

Overall, the uncertainty in the best fit density and radiation field strength for each galaxy typically cause the corresponding CO-to-H$_{2}$ and C-to-H$_{2}$ conversion factors to vary by no more than a factor of 2--3, whilst the adopted uncertainty in the metallicity has a stronger influence, producing a factor of 4 or higher variation in $X_{\rmn{CO}}$ and $X_{\text{C\,{\sc i}}}$. Conversion factors for the higher transition lines of CO are found to be more susceptible to the uncertainties in the physical parameters. The conversion factors derived for the SMC suffer the greatest level of uncertainty, a consequence of the poorer quality of fit that was obtained for this galaxy.

The essence of the X-factor method is that a single measurement of CO intensity -- representing a wide range of physical conditions in a region of space -- can be used to estimate the mass of gas in that region. Our method predicts the most probable values for the physical conditions within the region considered, and hence the most probable value of the CO (or other) emission. We believe, therefore, that our method is consistent with the X-factor method and that our derived conversion factor values are appropriate to use, rather than other values from the possible range of $X_{\rmn{CO}}$ obtained as above.

In future work, the range of physical parameters over which each specified conversion factor remains valid will be examined in more detail, providing limits on their applicability. The emission lines of other species might also prove to be better tracers of the molecular mass under certain conditions (e.g.~HCN for dense gas); this prospect will also be considered in a future study. It is hoped that, taken together, these conversion factors will provide a firm basis upon which increasingly reliable molecular mass estimates can be made.


\section*{Acknowledgments}
We wish to thank Profs Mike Barlow and Tom Millar for helpful comments and discussions, and the referee for constructive comments which helped to improve an earlier draft of this paper. TAB is supported by a PPARC studentship. SV acknowledges individual financial support from a PPARC Advanced Fellowship.


\bsp

\label{lastpage}

\end{document}